\documentclass{elsart}

\usepackage{icarus}

\usepackage{natbib}
\usepackage{amssymb}
\usepackage{graphicx}

\bibpunct{(}{)}{;}{a}{,}{,}

\newcommand{\rn}{$^{15}$N/$^{14}$N}

\begin{document}
\begin{frontmatter}



\title{New constraints on the delivery of cometary water and nitrogen
to Earth from the \rn\ isotopic ratio}


\author[inst1,inst2]{Damien Hutsem\'ekers}, 
\author[inst1,inst3]{Jean Manfroid},
\author[inst1,inst4]{Emmanu\"el Jehin}, and
\author[inst1]{Claude Arpigny}

\address[inst1]{Institut d'Astrophysique et de G\'eophysique,
                Universit\'e de Li\`ege, All\'ee du 6 ao\^ut 17, 
                B-4000 Li\`ege (Belgium)}
\address[inst2]{Senior Research Associate FNRS}
\address[inst3]{Research Director FNRS}
\address[inst4]{Research Associate FNRS}



%
%
%
%
%


\end{frontmatter}



\begin{flushleft}
\vspace{1cm}
Number of pages: \pageref{lastpage} \\
Number of tables: 0 \\
Number of figures: 1 \\
\end{flushleft}


\begin{pagetwo}{The delivery of cometary water and nitrogen to Earth}

D. Hutsem\'ekers \\
Institut d'Astrophysique et de G\'eophysique\\
Universit\'e de Li\`ege\\
All\'ee du 6 ao\^ut 17\\
B-4000 Li\`ege, Belgium\\
\\
Email: hutsemekers@astro.ulg.ac.be\\
Phone: +32 4 366 9760 \\
Fax: +32 4 366 9746

\end{pagetwo}

\begin{abstract}

New independent constraints on the amount of water delivered to Earth
by comets are derived using the \rn\ isotopic ratio, measured to be
roughly twice as high  in cometary CN and HCN as in the present
Earth.  Under reasonable assumptions, we find that no more than a few
percent of Earth's water can be attributed to comets, in agreement
with the constraints derived from D/H.  Our results also suggest that
a significant part of Earth's atmospheric nitrogen might come from
comets.  Since the \rn\ isotopic ratio is not different in Oort-cloud
and Kuiper-belt comets, our estimates apply to the contribution of
both types of objects.

\end{abstract}



\section{Introduction}

The origin of water on Earth is still puzzling \citep[e.g. the reviews
by][and references therein]{owe98,rob01,dra05,jav05,mar06}.  At one
end of the models, the high temperature in the inner accretion disk
hampered hydrous phases to exist so that external sources were needed,
the so-called late veneer of comets, asteroids and/or meteorites
suggested by lunar cratering \citep[e.g.][]{gom05}.  At the other end,
Earth accreted hydrous silicate phases, or grains having adsorbed
water from the solar nebula. The D/H isotopic ratio plays a key role
in constraining the models. The comparable D/H ratio of Earth's oceans
and carbonaceous chondrites (the only ones to contain enough water)
points towards a meteoritic veneer, while the cometary D/H ratio,
twice as high as the terrestrial one, would require a mixture of
cometary water with a roughly equal amount of primitive indigeneous
D-poor terrestrial water.  While the abundance of noble gases in the
Earth's atmosphere suggests that only small amounts of cometary
volatiles were brought to Earth, the $^{187}$Os/$^{188}$Os ratio
apparently rules out the carbonaceous chondrites as the main component
of the veneer \citep[][and references therein]{dra05}. However, most
of these arguments have shortcomings \citep[see discussions
by][]{dra05,jav05,mar06,dau03,gen08}. In particular, D/H has only been
measured in Oort-Cloud comets and it is not excluded that other types
of comets --or other reservoirs of hydrogen in comets-- exhibit
different D/H ratios. Additional constraints are therefore mandatory.

Recently, we found that the isotopic ratio \rn\ is consistently twice
as high in cometary CN and HCN as in the Earth's atmosphere and
surface \citep{arp03,boc08,man09}.  Since any delivery of water by
comets would necessarily be accompanied by a delivery of nitrogen, the
observed difference between terrestrial and cometary \rn\ can induce a
measurable isotopic shift and provide independent constraints on the
amount of water and nitrogen delivered by comets to Earth's oceans and
atmosphere, as detailed below. The interest of considering the
nitrogen isotopic ratio has been recently emphasized by \citet{mar06}.

\clearpage

\section{Constraints from D/H : framework and previous results}

We assume that cometary H$_2$O and HDO add to a primitive mixture on
Earth to give the amounts presently measured in Earth's water. We have
then
\begin{eqnarray}
n_p ({\rm H_2O}) + \ n_c ({\rm H_2O}) & = & n_t ({\rm H_2O}) \nonumber \\ 
n_p ({\rm HDO})  + \ n_c ({\rm HDO})  & = & n_t ({\rm HDO}) 
\end{eqnarray}
where $n$ is the abundance in number, the suffix $p$, $c$, and $t$
respectively referring to primitive Earth, cometary and terrestrial
(present Earth) values. From these equations we derive
\begin{eqnarray}
\frac{n_c ({\rm H_2O})}{n_t ({\rm H_2O})} = \frac{({\rm D/H})_p-({\rm D/H})_t}{({\rm D/H})_p-({\rm D/H})_c}
\end{eqnarray}
which provides the proportion of Earth's water due to comets as a
function of the isotopic ratio D/H in water.

The present Earth D/H ratio is fairly well known and measured to be
(D/H)$_t$ = 1.49$\pm$0.03 10$^{-4}$ for water stored at the Earth's
surface, mainly in the oceans \citep{lec98}. For cometary water we
have (D/H)$_c$ = 3.1$\pm$0.3 10$^{-4}$ from in situ measurements of
comet 1P/Halley \citep{bal95,ebe95} and from remote observations of
three Oort-Cloud comets: C/1996 B2 (Hyakutake) \citep{boc98}, C/1995
O1 (Hale-Bopp) \citep{mei98a} and C/2002 T7 (LINEAR)
\citep{hut08}. The primitive Earth D/H ratio is not known. At one
extreme, we have the protosolar value of the huge H$_2$ reservoir
measured from DH/H$_2$ in Jupiter: (D/H)$_p \geq$ 0.26 10$^{-4}$
\citep{mah98}. At the other end, the D/H ratio estimated to be
representative of the deep mantle (D/H)$_p \leq$ 1.36 10$^{-4}$
\citep{del91,dau00}. In the case that water from the warm, inner solar
nebula is adsorbed by the rocky grains that formed the bulk of the
Earth, intermediate values (D/H)$_p \simeq$ 0.8--1.0 10$^{-4}$ are
likely to better characterize the primitive Earth D/H ratio
\citep{owe98}. With these values, the proportion of Earth's water due
to comets varies between 50\% and 10\% (Eq.~2), with plausible values
up to 30\% \citep{ebe95,boc98,owe98}. More detailed modelling,
combining the D/H ratios with lunar cratering records and terrestrial
mantle siderophiles, suggests that the contribution of comets to
Earth's water is smaller than 10\% \citep{dau00}.  Moreover, should a
significant fraction of cometary material possess an even higher D/H
as measured in HCN \citep{mei98b} and assuming that this material can
be recycled at the Earth's surface, even smaller cometary
contributions would be expected.

\section{New constraints from \rn }

Assuming similarly that the nitrogen isotopic ratio of the present
Earth results from the combination of primitive Earth and cometary
nitrogen mixtures, we have
\begin{eqnarray}
n_p (^{14}{\rm N})  + \ n_c (^{14}{\rm N})  & = & n_t (^{14}{\rm N}) \nonumber \\ 
n_p (^{15}{\rm N})  + \ n_c (^{15}{\rm N})  & = & n_t (^{15}{\rm N})  \,\, .
\end{eqnarray}
and 
\begin{eqnarray}
\frac{n_{c} (^{14}{\rm N})}{n_t (^{14}{\rm N})} = 
\frac{(^{15}{\rm N}/^{14}{\rm N})_p - (^{15}{\rm N}/^{14}{\rm N})_t}{(^{15}{\rm N}/^{14}{\rm N})_p 
- (^{15}{\rm N}/^{14}{\rm N})_{c}} \,\, .
\end{eqnarray}

Since nitrogen is mostly in the form $^{14}$N, we may write
\begin{eqnarray} 
\frac{n_c ({\rm H_2O})}{n_t ({\rm H_2O)}} \ = \ \frac{n_{c} (^{14}{\rm N})}{n_t (^{14}{\rm N})}\ \times \ \frac{n_t ({\rm N})}{n_t ({\rm H_2O)}} \ \times \  \left( \frac{n_{c} ({\rm N})}{n_{c} ({\rm H_2O)}} \right) ^{-1}  \ \,\, ,
\end{eqnarray}
with the constraint
\begin{eqnarray} 
\frac{n_c ({\rm H_2O})}{n_t ({\rm H_2O)}} \ \leq \  \frac{n_t ({\rm N})}{n_t ({\rm H_2O)}}  \ \times \  \left( \frac{n_{c} ({\rm N})}{n_{c} ({\rm H_2O)}} \right) ^{-1} 
\end{eqnarray}
which comes from the fact that the total amount of nitrogen delivered
by comets to Earth cannot exceed the amount of nitrogen presently
measured in Earth's atmosphere and surface, i.e.  $n_{c} (^{14}{\rm
N}) / n_t (^{14}{\rm N}) \leq 1$.  Nitrogen recycling at Earth's
surface is implicitly assumed as well as the fact that the
composition of the atmosphere has not significantly changed since the
late veneer \citep{tol98}.

The amount of water and nitrogen at the surface of the present Earth
(mostly in the oceans and in the atmosphere, respectively) is
relatively well known. The water inventory by \citet{lec98} \citep[see
also][]{dau00} gives 1.7 10$^{21}$ kg of water at the Earth's surface
(oceans, ice sheets, organic matter, metamorphic rocks, shales,
sandstones, continental carbonates, evaporites, marine clays and
marine carbonates). The nitrogen inventory at the Earth's surface
(atmosphere, sedimentary rocks, crustal igneous rocks) gives 5.0
10$^{18}$ kg of N$_2$ \citep{zha93}. Thus $n_t ({\rm N}) / n_t ({\rm
H_2O)}$ $\simeq$ 3.8~10$^{-3}$.

The abundance of nitrogen in comets is poorly known. Because N$_2$ is
not detected, the nitrogen inventory in cometary ices mainly relies on
the measurement of the NH$_3$ and HCN volatiles.  From the abundances
given by \citet{boc04}, we compute ${n_c({\rm N}) / n_c ({\rm H_2O)}}
\gtrsim$ 10$^{-2}$ adding the contributions of all observed N-bearing
molecules and considering that contributions from other molecules are
possible.  Based on in-situ measurements by the Giotto spacecraft,
\citet{enc91} estimated the total N/O abundance in 1P/Halley: N/O =
0.027$\pm$0.009, including both the gas and the dust
components. Assuming that at most 60\% of O is in H$_2$O
\citep{enc91}, ${n_c({\rm N}) / n_c ({\rm H_2O)}} \gtrsim$ 3
10$^{-2}$. Further modelling by \citet{gre98} and \citet{hue02}
suggests that ${n_c({\rm N}) / n_c ({\rm H_2O)}}$ may be as high as
7~10$^{-2}$, nitrogen being still depleted in comets by a factor 3
with respect to solar abundances. Under the hypothesis that the
nitrogen abundance in 1P/Halley is representative of comets, we adopt
the lower limit ${n_c({\rm N}) / n_c ({\rm H_2O)}} \gtrsim$ 3
10$^{-2}$ from which we derive $n_c ({\rm H_2O})/n_t ({\rm H_2O)}$
$\lesssim$ 13\% (Eq.~6).

The present Earth \rn\ ratio is accurately measured from N$_2$ in the
atmosphere: (\rn)$_t$ = 3.676 $10^{-3}$ \citep{jun58}.  This is
comparable to \rn\ = 3.64 10$^{-3}$ which characterizes the present
mantle \citep{bec03}.  The primitive Earth \rn\ is not known. It may
be close to \rn\ $\simeq$ 3.55 10$^{-3}$, the value measured in
enstatite chondrites thought to have released significant amounts of
nitrogen during and shortly after the accretion phase \citep{jav86}.
On the other hand, a massive atmosphere could have been captured
directly from the solar nebula \citep{pep06,gen08} where \rn\ $\simeq$
2.3 10$^{-3}$ \citep[][this value is estimated from the atmosphere of
Jupiter assumed to be a proxy of the N$_2$ reservoir in the solar
system.]{owe01,mei07}.  A large part of the early atmosphere might
have been lost during the post-accretion phase (much before the late
veneer), possibly resulting in a small shift of the primitive \rn\
ratio due to atmospheric escape \citep{tol98}. Given these
uncertainties we consider in the following both (\rn)$_p$ = 2.3
10$^{-3}$ and (\rn)$_p$ = 3.55 10$^{-3}$. Intermediate values are
possible, as well as values closer to the present terrestrial value if
fractionation by atmospheric escape has been significant.

\rn\ has been measured remotely in about twenty comets using
optical-UV spectroscopy of CN, and in three comets on the basis of
radio observations of HCN. Recent studies show that optical and radio
measurements do agree within uncertainties \citep{boc08}.  The
nitrogen isotopic ratio is remarkably similar from comet to comet and
clusters around the ``anomalous'' \rn\ = 6.8$\pm$0.3 10$^{-3}$ which
is twice as high as the terrestrial value \citep{man09}. On the
other hand, the analysis of comet 81P/Wild2 grains returned by
Stardust suggests bulk \rn\ ratios comparable to the terrestrial and
chondritic values \citep{mck06}. We therefore compute an average
cometary ratio according to
\begin{eqnarray}
(^{15}{\rm N}/^{14}{\rm N})_{c} = 
\frac{n_{ca} ({\rm N})}{n_c ({\rm N})} (^{15}{\rm N}/^{14}{\rm N})_{ca}
+(1 - \frac{n_{ca} ({\rm N})}{n_c ({\rm N})})  (^{15}{\rm N}/^{14}{\rm N})_{cd}
\end{eqnarray}
where $(^{15}{\rm N}/^{14}{\rm N})_{ca}$ = 6.8 10$^{-3}$, $(^{15}{\rm
N}/^{14}{\rm N})_{cd}$ = 3.55 10$^{-3}$, and $n_{ca} ({\rm N})$ is the
abundance of the isotopically anomalous nitrogen.  

In Fig.~1 we illustrate the proportion of terrestrial nitrogen
delivered by comets, $n_c ({\rm N})/n_t ({\rm N)}$, as a function of
the relative abundance of anomalous nitrogen ${n_{ca}({\rm N}) / n_{c}
({\rm H_2O)}}$ for various cases of interest (Eqs. 4 and 7). Fig.~1
shows that the nitrogen isotopic shift between primitive and present
Earth can be easily accounted for by the addition of a cometary
component thanks to the presence, even in small quantities, of
isotopically anomalous nitrogen in comets.  The proportion of
terrestrial water delivered by comets can be computed from Fig.~1 and
Eq.~5, i.e. $n_c ({\rm H_2O})/n_t ({\rm H_2O)}$ $\lesssim$
0.13$\times$$n_c ({\rm N})/n_t ({\rm N)}$ for ${n_c({\rm N}) / n_c
({\rm H_2O)}} \gtrsim$ 3 10$^{-2}$.

Assuming that the carrier of the isotopically anomalous nitrogen is
only HCN, we have ${n_{ca}({\rm N}) / n_{c} ({\rm H_2O)}} \simeq$ 2
10$^{-3}$ using the abundances of \citet{boc04}.  Although $^{15}{\rm
N}/^{14}{\rm N}$ has not yet been measured in cometary ammonia,
fractionation mechanisms predict that $^{15}$N must be enhanced in
NH$_3$ before being transferred to HCN compounds \citep{cha02,rod08}.
If NH$_3$ is also included in the carriers of the anomalous \rn\
ratio, we have ${n_{ca}({\rm N}) / n_c ({\rm H_2O)}} \simeq$ 8
10$^{-3}$ using 6 10$^{-3}$ for the mean abundance of NH$_3$ relative
to H$_2$O \citep{boc04}. In this case between 15\% and 65\% of Earth's
nitrogen might have been delivered by comets (Fig.~1).  We then
estimate the upper limit $n_{c} ({\rm H_2O})/n_t ({\rm H_2O)}$
$\lesssim$ 9\%, which can be as low as $n_{c} ({\rm H_2O})/n_t ({\rm
H_2O)}$ $\lesssim$ 2\% if the primitive Earth \rn\ was close to 3.55
10$^{-3}$. Given that the abundance of nitrogen in comets is probably
larger than ${n_c({\rm N}) / n_c ({\rm H_2O)}}$ = 3 10$^{-2}$, and
that the refractory component may also contain isotopically anomalous
nitrogen, the contribution of comets to Earth's water is probably
roughly twice as small as the upper limit we just derived, i.e. not
larger than a few percent.

These constraints are more stringent than the values derived from D/H
using Eq.~2. Moreover, as soon as $n_{c} ({\rm H_2O})/n_t ({\rm
H_2O)}$ $\lesssim$ 7\%, the amount of water delivered by comets is not
sufficient to explain the D/H isotopic shift in terrestrial oceans
from the primitive value (Eq.~2) so that additional sources of water
are needed.  Our estimates agree with the constraints derived from
recent dynamical models of the solar system \citep{mor00}, and from
mass-balance models based on D/H \citep{dau00} or based on noble
metals and gases \citep{dau02}.

As far as the atmosphere is concerned, \citet{mar07} showed that a
cometary contribution fitting the abundances of noble gases in the
Earth's atmosphere would deliver $\sim$ 6\% of the atmospheric
nitrogen. Given the uncertainties on the abundances of noble gases in
comets \citep{boc04}, this is compatible with our estimates provided
that ${n_{ca}({\rm N}) / n_{c} ({\rm H_2O)}} \simeq$ 10$^{-2}$ and
(\rn)$_p$ $\simeq$ 3.55 10$^{-3}$ in Fig.~1.  Early fractionation due
to atmospheric escape \citep{tol98} might also have slightly increased
(\rn)$_p$ providing an even better agreement.

It is important to note that the anomalous \rn\ isotopic ratio was
measured to be identical in Oort-cloud and Kuiper-belt comets
\citep{hut05,man09} so that our estimates apply to the contributions
of both types of comets.

\section{Conclusions}

Thanks to the fact that the \rn\ ratio measured  in cometary CN
and HCN is significantly different from the ratio measured on Earth,
and since any delivery of nitrogen from comets to Earth necessarily
accompanied that of water, we put independent constraints on the
amount of Earth's water possibly due to comets.  Under reasonable
assumptions, we find that no more than a few percent of Earth's water
can be attributed to comets.  This is consistent with the constraints
derived from D/H using various models \citep{mor00,dau00,dau02}.
Since the \rn\ isotope ratio is not different in Oort-cloud and
Kuiper-belt comets, our estimates apply to both types of objects.  Our
results also suggest that a significant part of Earth's nitrogen might
come from comets, supporting the idea of a dual origin of the Earth's
atmosphere \citep{owe98,dau03,mar07}.  A critical measurement to
further constrain these quantities would be the determination of the
\rn\ ratio in cometary NH$_3$, expected to be either anomalous or
terrestrial.  Although more detailed modelling is required for more
quantitative estimates, our results demonstrate the interest of
considering the \rn\ ratio to evaluate the contribution of comets to
the late bombardment of the Earth.

\ack
We are grateful to Dominique Bockel\'ee-Morvan, Bernard Marty
and anonymous referees for comments which helped to significantly
improve the paper.

\label{lastpage}

\newpage

\begin{figure}[t]
\begin{center}
\resizebox{\hsize}{!}{\includegraphics*{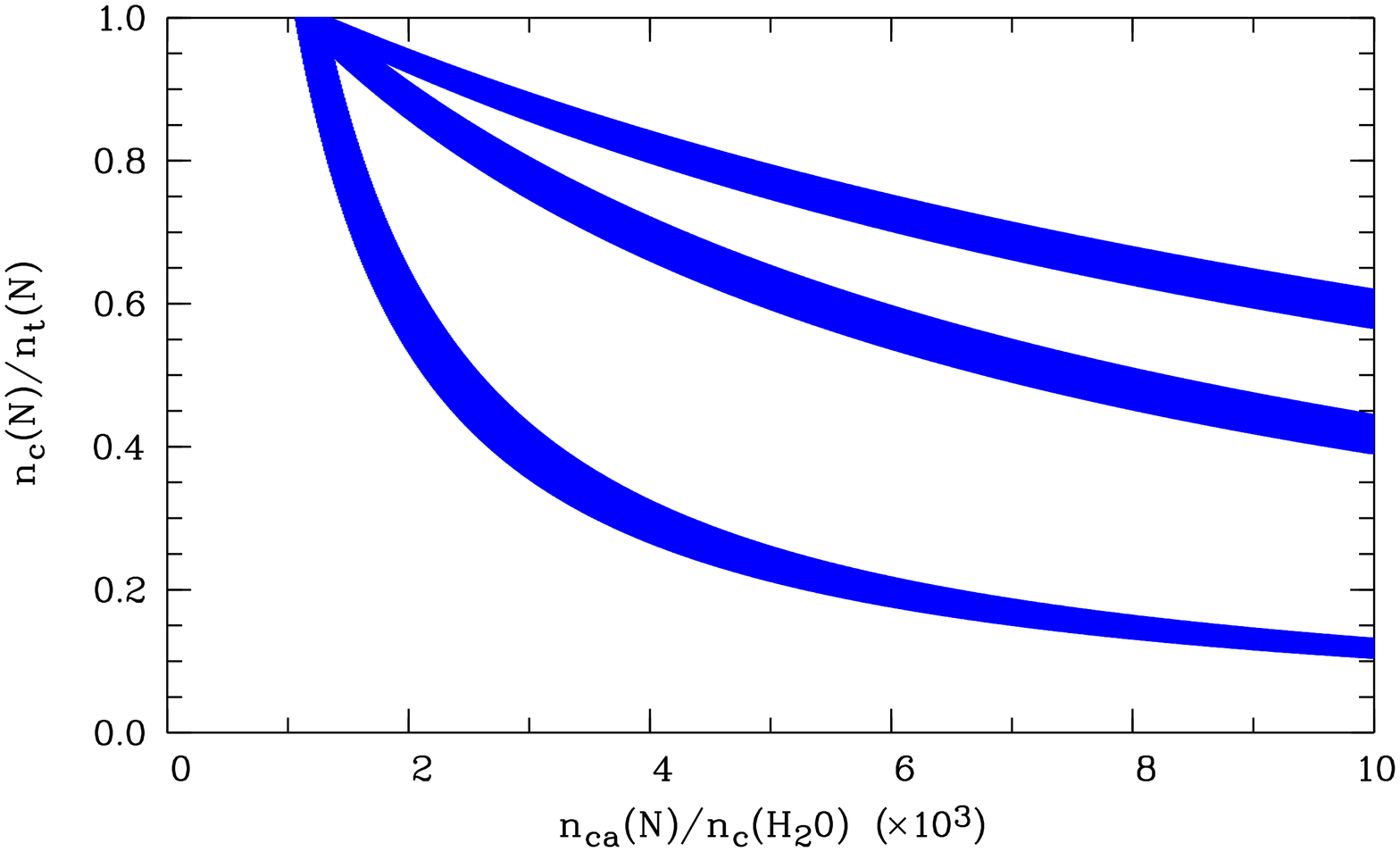}}
\caption{The proportion of terrestrial nitrogen due to comets $n_c
({\rm N})/n_t ({\rm N)}$ as a function of the abundance of
isotopically anomalous nitrogen in comets ${n_{ca}({\rm N}) / n_{c}
({\rm H_2O)}}$.  The three curves are computed from Eqs. 4 and 7 with
${n_{c}({\rm N}) / n_{c} ({\rm H_2O)}}$ = 3 10$^{-2}$.  The proportion
of terrestrial water due to comets is given by $n_c ({\rm H_2O})/n_t
({\rm H_2O)}$ $\lesssim$ 0.13$\times$$n_c ({\rm N})/n_t ({\rm N)}$
(Eq.~5).  Each curve corresponds to a different value of the primitive
Earth nitrogen isotopic ratio : (\rn )$_p$ = 2.3 10$^{-3}$, 3.0
10$^{-3}$ and 3.55 10$^{-3}$ from top to bottom.  The width of the
curves accounts for the uncertainty on the measured cometary (\rn
)$_{ca}$.}
\label{fig:fig1}
\end{center}
\end{figure}

\end{document}